\documentclass{article}

\usepackage{arxiv}

\usepackage[utf8]{inputenc} % allow utf-8 input
\usepackage[T1]{fontenc}    % use 8-bit T1 fonts
\usepackage{hyperref}       % hyperlinks
\usepackage{url}            % simple URL typesetting
\usepackage{booktabs}       % professional-quality tables
\usepackage{amsfonts}       % blackboard math symbols
\usepackage{nicefrac}       % compact symbols for 1/2, etc.
\usepackage{microtype}      % microtypography
\usepackage{lipsum}		% Can be removed after putting your text content
\usepackage{graphicx}
\usepackage{doi}

\title{2D and 3D CNN-Based Fusion approach for COVID-19 Severity Prediction From 3D CT-Scans}

\author{ \href{https://sciprofiles.com/profile/FaresBougourzi}{\includegraphics[scale=0.06]{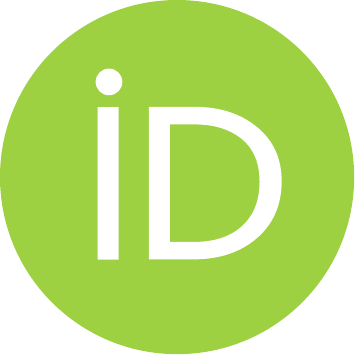}\hspace{1mm}Fares BOUGOURZI}
\thanks{.} \\
University Paris-Est Creteil, Laboratoire LISSI, 94400 \\Vitry sur Seine, Paris, France \\
	\texttt{faresbougourzi@gmail.com} \\
	\And
	\href{https://orcid.org/0000-0000-0000-0000}{\includegraphics[scale=0.06]{orcid.pdf}\hspace{1mm}Fadi DORNAIKA} \\
	University of the Basque Country UPV/EHU,\\
	San Sebastian, SPAIN; IKERBASQUE, Basque \\ Foundation for Science, Bilbao, SPAIN \\
	\texttt{fadi.dornaika@ehu.eus} \\  
	\And
	\href{https://orcid.org/0000-0000-0000-0000}{\includegraphics[scale=0.06]{orcid.pdf}\hspace{1mm}Amir Nakib} \\
	University Paris-Est Creteil, Laboratoire LISSI,\\ 94400 Vitry sur Seine, Paris, France \\
	\texttt{nakib@u-pec.fr} \\ 
	\And
	\href{https://orcid.org/0000-0000-0000-0000}{\includegraphics[scale=0.06]{orcid.pdf}\hspace{1mm}Cosimo Distante} \\
	Institute of Applied Sciences and Intelligent\\ Systems, National Research Council of Italy,\\ 73100 Lecce, Italy \\ Department of Innovation Engineering, University \\of Salento, 73100 Lecce, Italy \\
	\texttt{cosimo.distante@cnr.it} \\	
	\And
	\href{https://orcid.org/0000-0000-0000-0000}{\includegraphics[scale=0.06]{orcid.pdf}\hspace{1mm}Abdelmalik Taleb-Ahmed} \\
	IEMN UMR CNRS 8520, Université \\Polytechnique Hauts de France, UPHF\\
	\texttt{Abdelmalik.Taleb-Ahmed@uphf.fr} \\
}

\renewcommand{\shorttitle}{\textit{arXiv} Template}

\hypersetup{
pdftitle={A template for the arxiv style},
pdfsubject={q-bio.NC, q-bio.QM},
pdfauthor={David S.~Hippocampus, Elias D.~Striatum},
pdfkeywords={First keyword, Second keyword, More},
}

\begin{document}
\maketitle

\begin{abstract}
Since the appearance of Covid-19 in late 2019, Covid-19 has become an active research topic for the artificial intelligence (AI) community. One of the most interesting AI topics is Covid-19 analysis of medical imaging. CT-scan imaging is the most informative tool about this disease. 

This work is part of the 3nd COV19D competition for Covid-19 Severity Prediction. In order to deal with the big gap between the validation and test results that were shown in the previous version of this competition, we proposed to combine the prediction of 2D and 3D CNN predictions. For the 2D CNN approach, we propose 2B-InceptResnet architecture which consists of two paths for segmented lungs and infection of all slices of the input CT-scan, respectively. Each path consists of ConvLayer and Inception-ResNet pretrained model on ImageNet.
For the 3D CNN approach, we propose  hybrid-DeCoVNet architecture which consists of four blocks: Stem, four 3D-ResNet layers, Classification Head and Decision layer.

Our proposed approaches outperformed the baseline approach in the validation data of the 3nd COV19D competition for Covid-19 Severity Prediction by 36\%.

\end{abstract}

% keywords can be removed
\keywords{Covid-19 \and Deep Leaning \and CNNs \and Recognition \and Severity }

\section{Introduction}
Since the appearance of the Covid-19 pandemic in the late of 2019, reverse transcription-polymerase chain reaction (RT-PCR) has been considered as the golden standards for Covid-19 Detection. However, the RT-PCR test has many drawbacks including inadequate supply of RT-PCR kits, time-consuming consumption, and considerable false negative results \cite{jin_rapid_2020, wu_jcs_2021, kucirka_variation_2020}. To deal with this limitations, Medical Imaging modalities have been widly used as supporting tools. These imaging modalities include X-rays and CT-scans \cite{vantaggiato2021covid,bougourzi_recognition_2021}. Indeed, CT-scans are not only used to detect Covid-19 infected cases, but they could be used to follow up the state of the patient and predicting the disease severity \cite{bougourzi_per-covid-19_2021, bougourzi_challenge_2021}.

In the last decade, Deep Learning methods have become dominant in most of the computer vision tasks and they have achieved high performance  compared to traditional methods \cite{bougourzi_fusing_2020, bougourzi_deep_2022}. However, the main drawback of deep learning, especially the CNN architecture is the need of huge labelled data, which is hard to be obtained in medical domains \cite{bougourzi_per-covid-19_2021}. On the other hand, most of the proposed CNN architectures adopted for Static images (single image input) \cite{bougourzi_challenge_2021}.

In this work, we proposed to combine the predictions of trained 3D and 2D CNN architectures for Covid-19 Severity Prediction from 3D CT-scans as part  of 3nd COV19D Competition. In order to deal with the big gap between the validation and test results that were shown in the previous version of this competition, we proposed to combine the prediction of 2D and 3D CNN predictions. For the 2D CNN approach, we propose 2B-InceptResnet architecture which consists of two paths for segmented lungs and infection of all slices of the input CT-scan, respectively. Each path consists of ConvLayer and Inception-ResNet pretrained model on ImageNet. For the 3D CNN approach, we propose  Hybrid-DeCoVNet architecture which consists of four blocks: Stem, four 3D-ResNet layers, Classification Head and Decision layer. 
The main contributions of this paper can be summarized as follows: 

\begin{itemize} 
\item We propose a combination trained 3D and 2D CNN architectures for Covid-19 severity prediction.

\item We propose 2B-InceptResnet architecture which consists of two paths for segmented lungs and infection, where this architecture aims to exploit pretrained weights on Imagenet.

\item We propose a Hybrid-DeCoVNet architecture for Covid-19 Severity Prediction.

\item The codes used and the pre-trained models are publicly available in \url{https://github.com/faresbougourzi/3rd-COV19D-Competition}. ( Last accessed on March, 15{$^{th}$} 2023).

\end{itemize}

This paper is organized in following way:  Section \ref{S:1} describes our proposed approaches for Covid-19 Detection and Severity Detection. The experiments and results are described in Section \ref{S:2}. Finally, we concluded our paper in Section \ref{S:3}.

%%%%%%%%%%%%%%%%%%%%%%%%%%%%%%%%%%%%%
\section{Our Approaches}
\label{S:1}

In this paper, we summarize the obtained results COVID-19 Severity Detection Challenge, which is part of the 3rd COV19D Competition and held in conjunction with IEEE ICASSP 2023.

Table \ref{tab:oldcha} summarizes the results of the 2nd COV19D competition for Covid-19 Severity Prediction. As shown in this table, all of the proposed approaches performance considerably drop in the testing phase compared with the validation performance. The aim of this work is to reduce the overfitting and having more generalization ability. To this end, we trained models provided training and validation splits then we gathered all the data then we performed 5 Fold cross-validation. In the testing phase the ensemble of all models (trained on different data splitting) will be evaluated on the unseen testing data.

%%%%%%%%%%%%%%%%%%%%%%%%%%%%%%%%%%%%%%%%%%%%
\begin{table}[h]
 \caption{2nd COV19D competition for Covid-19 Severity Prediction Results. }
\label{tab:oldcha}
\centering
\begin{tabular}{|p{3cm}|p{2cm}|p{2cm}|}
\hline

\textbf{Team}  & \textbf{Validation} &\textbf{Test}  \\\hline
 FDVTS \cite{hou2023boosting}&77.03&  51.76 \\ \hline 
 Code 1055 \cite{bougourzi2023cnr}&68.18 &  51.48\\ \hline
CNR-IEMN \cite{kienzle2023covid}& 80.08& 47.11  \\ \hline 
MDAP \cite{turnbull2023using}& 75.46 & 46.00 \\ \hline 
%MDAP \cite{tan2023two}&68.55  & 41.49 \\ \hline 
\end{tabular}
\end{table}
%%%%%%%%%%%%%%%%%%%%%%%%%%%%%%%%%%%%%%%%%%%%

%\cite{hou2023boosting,bougourzi2023cnr, kienzle2023covid,turnbull2023using}

%%%%%%%%%%%%%%%%%%%%%%%%%%%%%%%%%%%%%%%%%%%%
%%%%%%%%%%%%%%%%%%%%%%%%%%%%%%%%%%%%%%%%%%%%
\subsection{Data Preprocessing}
\label{sec:headings}

The aim of this phase is to remove the slices that do not show lungs at all and to detect the features inside the lung for the remaining slices. Since each 3D CT scans can have  multiple slices with no lung regions, we trained a ResneXt-50 model \cite{Resnext} to filter the non-lung slices and keep just the slices where the lung appears. To this end,  we manually labelled 20 CT-scans from the training data as lung and non-lung classes. Furthermore, three Covid-19 segmentation datasets are used: COVID-19 CT segmentation \cite{COVID-19-Dataset},  Segmentation dataset nr.2 \cite{COVID-19-Dataset}, and COVID-19- CT Seg dataset \cite{2021MedPh..48.1197M}. In the second preprocessing phase, we trained Attention-Unet \cite{oktay_attention_2018, bougourzi2022ilc}  architecture to segment the lungs and Covid-19 infection. The same datasets used for filtering the slices are used to train and validate the Attention-Unet architecture, where the three datasets are combined and randomly splited into train-val splits corresponding to 70-30\%, respectively.

%%%%%%%%%%%%%%%%%%%%%%%%%%%%%%%%%%%%%%%%%%%%
%%%%%%%%%%%%%%%%%%%%%%%%%%%%%%%%%%%%%%%%%%%%
\subsection{2D-based CNN architecture}
\label{sec:headings}
\begin{figure}
\centering
    \includegraphics[width = 7in, height = 4in]{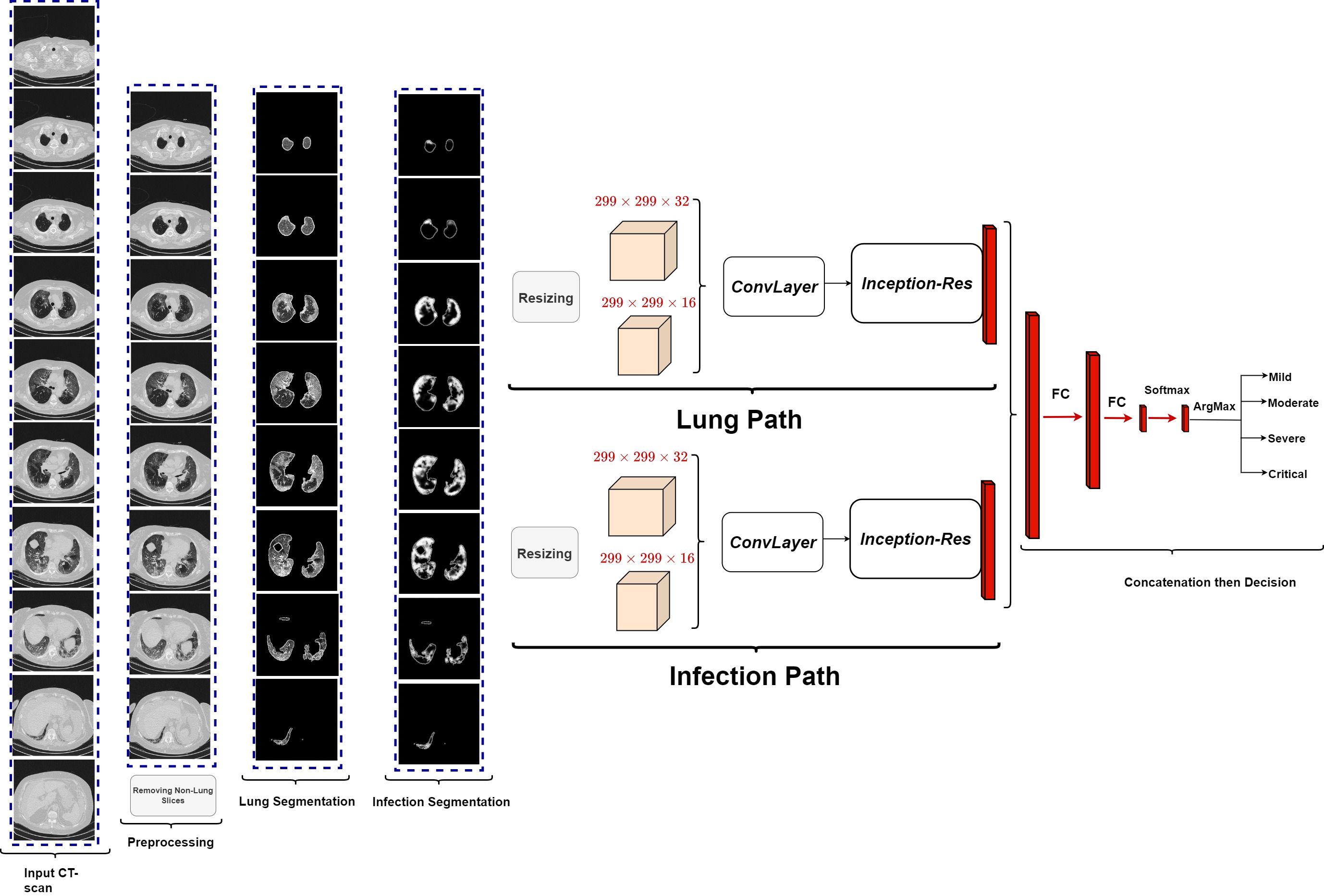} 
    \caption{Our proposed 2B-InceptResneth.}
    \label{fig:approach}
\end{figure}

\begin{figure}
    \includegraphics[width = 5in, height = 1in]{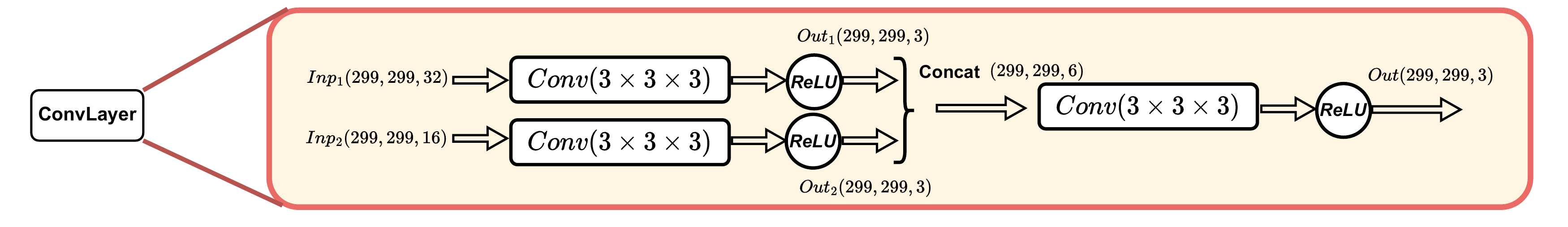} 
    \caption{Our proposed ConvLayer block.}
    \label{fig:approach2}
\end{figure}

Inspired by our proposed  CNR-IEMN approach \cite{kienzle2023covid}, that were proposed for the 2nd COV19D challenge, we propose two branches Inception-Resnet architecture (2B-InceptResnet) as shown in Figure \ref{fig:approach}. After segmenting both the lung and Covid-19 infection from the filtred slices, each segmentation types slices are concatenated then resized into two volumes which have the shapes $299\times299\times32$ and $299\times299\times16$, respectively. The aim of resizing each volume into two size is to reduce the information loss due to resizing operation and having two views to the same input. Both the lungs and infection volumes are feed into a ConvLayer, which is depicted in Figure \ref{fig:approach2}, then the resulting tensor is feed to Inception-Resnet architecture. The last deep features from each path are concatenated then passed to two FC layer to predict the input CT-scans severity. 

%%%%%%%%%%%%%%%%%%%%%%%%%%%%%%%%%%%%%%%%%%%%%%%%%%%%
%%%%%%%%%%%%%%%%%%%%%%%%%%%%%%%%%%%%%%%%%%%%%%%%%%[h]

%%%%%%%%%%%%%%%%%%%%%%%%%%%%%%%%%%%%%%%%%%%%%%%%%%%%
%%%%%%%%%%%%%%%%%%%%%%%%%%%%%%%%%%%%%%%%%%%%%%%%%%[h]
%%%%%%%%%%%%%%%%%%%%%%%%%%%%%%%%%%%%%%%%%%%%
%%%%%%%%%%%%%%%%%%%%%%%%%%%%%%%%%%%%%%%%%%%%
\subsection{3D-based CNN architecture}
\label{sec:headings}
\begin{figure}
\centering
    \includegraphics[width = 7in, height = 4in]{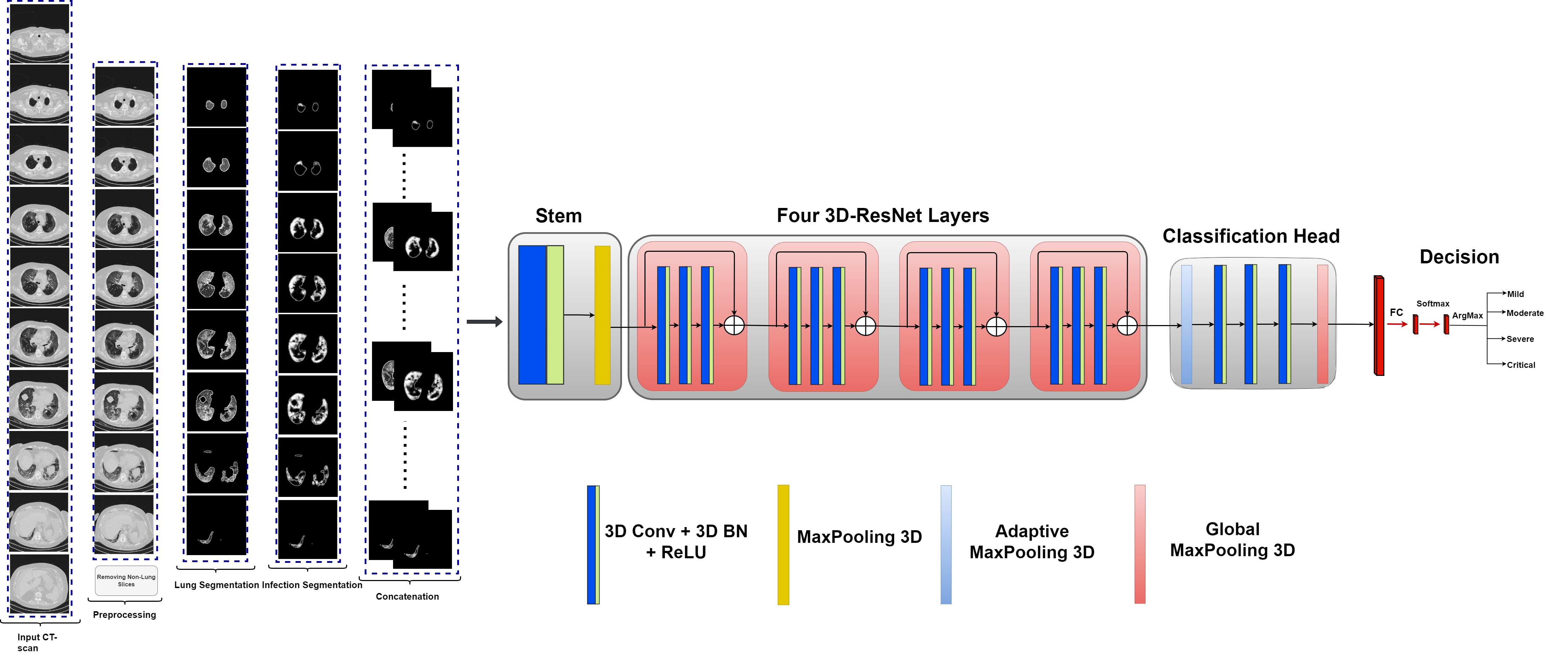} 
    \caption{Our proposed hybrid-DeCoVNet Approach.}
    \label{fig:approach3}
\end{figure}

Following the proposed DeCoVNet architecture in \cite{wang2020weakly}, we proposed the hybrid-DeCoVNet for Covid-19 Severity Prediction as shown in Figure \ref{fig:approach3}. The proposed Hybrid-DeCOVNet consists of four componets as show in Figure  \ref{fig:approach3}. First, the segmented lung and infection are concatenated to form two channels, then the concatenate segmented slices of each CT-scan are concatenated and resized into $224\times224\times2\times64$. The produced volume is feed into the first block of  hybrid-DeCoVNet, which is called Stem. The stem block is 3D convolutional layer with kernet (7, 7, 5) for the hight, width and depth, respectively. This convolutional kernel transforms the two input channels into 16 channels, and it is followed by Normalization Layer (BN) and ReLU activation function. The second block of Hybrid-DeCOVNet consists of four 3D-Resnet layers, that expands the channels into 64, 128, 256 and 512, respectively. The Classification Head consists of 3D Adaptive MaxPooling, three 3D concolutional layers, and 3D Global MaxPooling. The output of the Classification Head is flattened into one channel deep features map which is feed into Decision head that consists of one FC layer.

%%%%%%%%%%%%%%%%%%%%%%%%%%%%%%%%%%%%%%%%%%%%%%%%%%%%
%%%%%%%%%%%%%%%%%%%%%%%%%%%%%%%%%%%%%%%%%%%%%%%%%%[h]

%%%%%%%%%%%%%%%%%%%%%%%%%%%%%%%%%%%%%%%%%
\section{Experiments and Results}
\label{S:2}
\subsection{The COV19-CT-DB Database}
The COVID19-CT-Database (COV19-CT-DB) \cite{kollias2022ai, arsenos2022large, kollias2021mia, kollias2020deep, kollias2020transparent, kollias2018deep}  consists of chest CT scans that are annotated for the existence of COVID-19.

%%%%%%%%%%%%%%%%%%%%%%%%%%%%%%%%%%%%%%%%%%%%
\begin{table}[h]
 \caption{Data samples in each Severity Class }
\label{tab:datasev}
\centering
\begin{tabular}{|p{2.5cm}|p{2cm}|p{2cm}|}
\hline

\textbf{Severity Class}  & \textbf{Training} &\textbf{Validation}  \\\hline
1. mild& 133 &  31 \\ \hline 
2. moderate& 124& 20\\ \hline
3. severe& 166& 45  \\ \hline 
4. critical& 39& 5 \\ \hline 

\end{tabular}
\end{table}

%%%%%%%%%%%%%%%%%%%%%%%%%%%%%%%%%%%%%%%%%%%%

Further partitioning of the COVID-19 cases was based on the severity of COVID-19, which was given by four medical experts in a range of 1 to 4, with 4 denoting critical status. 
The training set contains a total of 462 3-D CT scans. The validation set consists of 101 3-D CT scans. The number of scans in each severity class  in these sets is shown in Table \ref{tab:datasev}.

\subsection{Experimental Setup}
For Deep Learning training and testing, we used the Pytorch \cite{paszke_pytorch_2019} Library with NVIDIA GPU Device GeForce TITAN RTX 24 GB. For the 2B-InceptResnet architecture training, the  used batch size consists of 16 CT-scan volumes and the architecture is trained for 40 epochs. The initial learning rate is 0.0001, which decreases by 0.1 after 15 epochs, followed by another 0.1 decrease after 30 epochs. The Hybrid-DeCOVNet is trained for 100 epochs since No pretrained weights are used. Both 2D and 3D CNN architectures are trained and tested on two scenarios: Train-Val and five folds cross validation. In the train and val scenario, we used the provided training and validation data by the 3rd COV19D organizers, in order to compare with the baseline results. To test the generalization ability of the proposed approach, we randomly splitted the combination of train and validation data in five folds.

\subsection{Results}
\label{sec:headings}

Table \ref{tab:Sevresult} and \ref{tab:Svfdcresult} summarize the results (F1 score) of detecting the severity of Covid-19 for train-val, and five folds cross validation scenarios, respectively. As described in section\ref{tab:Sevresult}, both the 2D and 3D approaches outperforms the baseline results by big margin, with small superiority for the 2D approach compared with the 3D approach. In addition, the combination of the two models predictions improves the results of the single architectures.
Similarly, Table \ref{tab:Svfdcresult} shows that the ensemble approach improves the results.
%%%%%%%%%%%%%%%%%%%%%%%%%%%%%%%%%%%%%%%%%%%%
\begin{table}[ht]
 \caption{Covid-19 Severity Results on the Train-Val scenario. }
\label{tab:Sevresult}
\centering
\begin{tabular}{|p{1cm}|p{5cm}|c|}
\hline

\textbf{Model}  &\textbf{Architecture}  & \textbf{Macro F1-Score}   \\\hline
- &Baseline& 38.00  \\ \hline 
1&3D-DeconvNet &68.40 \\ \hline
2&TwoB-InceptRes (Base) & 72.96\\ \hline 
3&Ensemble (2D + 3D) & 74.16\\ \hline

\end{tabular}
\end{table}

%%%%%%%%%%%%%%%%%%%%%%%%%%%%%%%%%%%%%%%%%%%%

%%%%%%%%%%%%%%%%%%%%%%%%%%%%%%%%%%%%%%%%%%%%
\begin{table}[ht]
 \caption{Covid-19 Severity Results on the five folds cross validation scenario. }
\label{tab:Svfdcresult}
\centering
\begin{tabular}{|p{1cm}|p{5cm}|c|c|c|c|c|}
\hline

\textbf{Model}  &\textbf{Architecture}  & \textbf{Fold1}& \textbf{Fold2}& \textbf{Fold3}& \textbf{Fold4}& \textbf{Fold5}   \\\hline
 
1&3D-DeconvNet & 71.26& 65.07&69.50 &67.23&65.03\\ \hline
2&TwoB-InceptRes (Base) & 73.28& 66.50& 73.20&67.01&64.91\\ \hline
3&Ensemble (2D + 3D) & 73.44& 67.45& 73.24&68.56&65.55\\ \hline

\end{tabular}
\end{table}

%%%%%%%%%%%%%%%%%%%%%%%%%%%%%%%%%%%%%%%%%%%%
%%%%%%%%%%%%%%%%%%%%%%%%%%%%%%%%%%%%%%%%%%%%
%%%%%%%%%%%%%%%%%%%%%%%%%%%%%%%%%%%%%%%%%%%%
\section{Conclusion}
\label{S:3}
In this work, we proposed an ensemble of 2D and 3D CNN-based approach for the 3nd COV19D competition for Covid-19 Severity Prediction. 
In order to deal with the big gap between the validation and test results that were shown in the previous version of this competition, we proposed to combine the prediction of 2D and 3D CNN predictions. For the 2D CNN approach, we propose 2B-InceptResnet architecture which consists of two paths for segmented lungs and infection of all slices of the input CT-scan, respectively. Each path consists of ConvLayer and Inception-ResNet pretrained model on ImageNet.
For the 3D CNN approach, we propose  hybrid-DeCoVNet architecture which consists of four blocks: Stem, four 3D-ResNet layers, Classification Head and Decision layer.

Our proposed approaches outperformed the baseline approach in the validation data of the 3nd COV19D competition for Covid-19 Severity Prediction by 36\%. In the testing phase, we will evaluate the ensemble performance of the trained 2D and 3D models using all splits.

%\bibliographystyle{unsrt}
%\bibliography{references}

\end{document}